# Phonons in Slow Motion: Dispersion Relations in Ultra-Thin Si Membranes

*John Cuffe[a,b], Emigdio Chavez[a,c], Andrey Shchepetov[d], P-Olivier Chapuis[a †], El Houssaine El Boudouti[e, f], Francesc Alzina[a], Timothy Kehoe[a], Jordi Gomis-Bresco[a], Damian Dudek[a ‡], Yan Pennec[e], Bahram Djafari-Rouhani[e], Mika Prunnila[d], Jouni Ahopelto[d] and Clivia. M. Sotomayor Torres∗ [a, c, g]*

[a] Catalan Institute of Nanotechnology, Campus UAB, 08193 Bellaterra (Barcelona), Spain
[b] Dept. of Physics, University College Cork, College Road, Cork, Ireland
[c] Dept. of Physics, Universitat Autonoma de Barcelona, 08193 Bellaterra (Barcelona), Spain
[d] VTT Technical Research Centre of Finland, PO Box 1000, 02044 VTT, Espoo, Finland
[e] Institut d'Electronique, de Microélectronique et de Nanotechnologie (IEMN), Université de Lille 1, France
[f] LDOM, Faculté des Sciences, Université Mohamed 1, Oujda, Morocco
[g] Institució Catalana de Recerca i Estudis Avançats (ICREA), 08010 Barcelona, Spain
[∗] Corresponding author. clivia.sotomayor@icn.cat

[†] *Present address: Centre de Thermique de Lyon (CETHIL)– CNRS – INSA Lyon, 9, rue de la Physique, Campus La Doua, 69621 Villeurbanne cedex, France*
[‡] *Present address: Deutsche Forschungsgemeinschaft (DFG), Kennedyallee, 53175 Bonn, Germany.*

We report the changes in dispersion relations of hypersonic acoustic phonons in free-standing silicon membranes as thin as ~ 8 nm. We observe a reduction of the phase and group velocities of the fundamental flexural mode by more than one order of magnitude compared to bulk values. The modification of the dispersion relation in nanostructures has important consequences for noise control in nano and micro-electromechanical systems (MEMS/NEMS) as well as opto-mechanical devices.

KEYWORDS: Confined phonons, slow phonons, ultrathin, Si membranes, dispersion relations, inelastic light scattering.

Advances in nanotechnology have opened exciting possibilities to control spectral distribution and propagation of phonons including, for example, optomechanical systems[1,2] and phononic crystals[2–5]. Recent experimental results[6–8] have suggested that slowing down phonons may have beneficial effects to improve the efficiency of thermoelectric materials, supporting earlier theoretical predictions[9–12]. On the other hand, as channel widths scale well below 20 nm, heat dissipation and phonon-limited electron mobility are major obstacles towards increasing device performance[13]. In all these cases, understanding phonon confinement at the nanoscale is essential[8,14].

There are at least two approaches to change dispersion relations. As with electrons, and later with photons and phonons, structures exhibiting a periodic contrast in a relevant property have led to tailoring dispersion relations and energy bands, thus engineering quantum wells in semiconductors[15], ultra-refraction in photonic crystals[16] and phonon cavities within phononic crystals[2,3,5]. The other approach relies on confinement, which also causes modifications of the dispersion relations[17,18]. These modifications affect phonon group velocities and are predicted to impact strongly on thermal conductivity[10,19,20] and charge carrier mobility[18,20–22]. A reduction in the characteristic dimension has been shown to limit the phonon mean free path[14] with a corresponding decrease in thermal conductivity. However, recent experimental work[6–8] suggested that the decrease in thermal conductivity of silicon phononic crystals could not be explained by this mean free path reduction alone, and further understanding of the effects of the phonon dispersion relation in nanostructures is required.

As mechanical eigenmodes of nanostructures, confined phonons also play a key role in ultra-sensitive mass sensors[23,24] and molecular-scale biosensing[25,26]. These investigations have resulted in breakthroughs such as reaching the quantum ground state of mechanical vibrations[27] and mass sensing with zeptogram resolution[23,24]. In the extreme sub- 10 nm regime, questions have arisen concerning the limits of validity of the continuum elasticity model and bulk elastic constants for nanoscale objects[28–30]. However, until now there has been a lack of experimental measurements relating to phonon propagation in the sub-50-nm regime, probably due to the challenging nature of the experiments with the desired resolution and the availability of samples with well-defined parameters[31].

The first direct studies of confined acoustic modes in ultra-thin free-standing silicon membranes were performed with Raman spectroscopy[32,33], a form of inelastic light scattering (ILS) spectroscopy. The increase in the mode frequencies and frequency spacing between them with decreasing membrane thickness was clearly observed. However, acoustic modes were only investigated at normal incidence, precluding observations of in-plane propagation and dispersion relations. While even earlier in-plane investigations were performed on unsupported 20 nm metal (Au) films[34], our interest in silicon is based on its well-known material properties, the unrivalled control of its nanofabrication and its still unchallenged supremacy as the material for micro-nano-electronic and MEMS devices.

Here, we use free-standing single-crystalline silicon membranes as model systems to test fundamental aspects and consequences of phonon confinement. Being unsupported, a true, two-dimensional geometry is obtained and the analysis is free from the effects of a substrate. The thickness values of the ultra-thin membranes investigated ranged from 7.8 ± 0.1 to 31.9 ± 0.2 nm. Membranes with thickness values of up to 400 nm were also investigated for comparison.

The dispersion relations of the confined phonons were measured by angle-resolved Brillouin scattering spectroscopy, another form of ILS spectroscopy proven to be a useful method to characterise acoustic properties[34,35] in a non-contact, non-destructive manner. The measurements were performed in backscattering configuration, with the incident wave vector, **k**, making an angle, $\theta$, to the surface normal of the sample as shown in Figure 1. Due to the in-plane momentum conservation[36], incident light of free-space wavevector $k_i = 2\pi/\lambda$ is inelastically scattered by phonons with a parallel wavevector component $q_{//}$ given by

$$q_{//} = 2k_i \sin\theta = \frac{4\pi}{\lambda}\sin\theta. \quad (1)$$

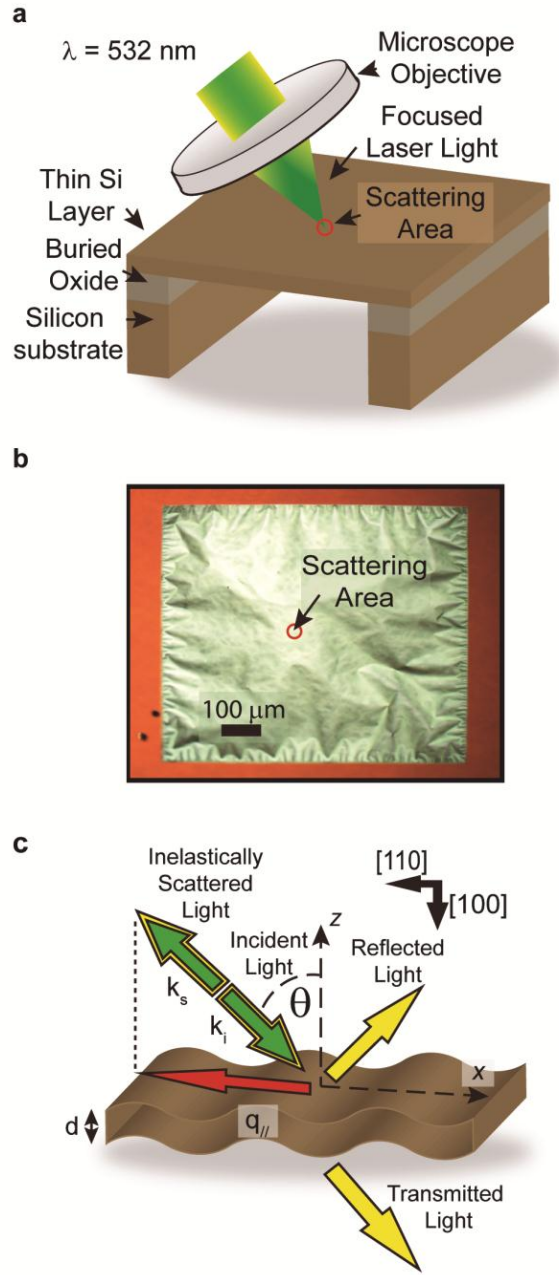

**Figure 1.** Membrane and scattering geometry. (a) The free-standing membrane samples were fabricated from (001) SOI wafers as explained in the Supplementary Information. The measurements were performed in backscattering configuration, with a 10x Olympus microscope objective used both to focus the incident light and collect the inelastically scattered light. (b) Optical microscope image of the 7.8 nm Si membrane. The observed wrinkles are a result of residual compressive strain from the buried oxide at the edges of the membrane. It was verified that these do not affect the measurement of the dispersion relation, as the scattering region is not significantly strained and is smaller than the characteristic length of the wrinkles. (c) Wavevector conservation rules, as explained in the text. The direction of the scattering wavevector, $q_{//}$, was kept in the [110] crystalline direction.

The dispersion relation ω(q$_{//}$) may thus be constructed by measuring the spectral components of the inelastically scattered light at well-defined angles.

In samples with a reduced scattering volume such as bulk opaque materials and nanometre-thin membranes, the surface ripple scattering mechanism is expected to be the dominant contribution to the inelastic light scattering[36]. At room temperature and above, such that $T \gg \hbar\omega/k_B$ ($k_B$ is the Boltzmann's constant, and $\hbar$ is the Plank's constant), the scattering intensity is related to the mean-squared amplitude of the out-of-plane surface displacement, $\langle|U_z|^2\rangle_{z=0}$[36],

$$I(\omega, q_{//}) \propto \langle|U_z|^2(\omega, q_{//})\rangle_{z=0} = \frac{k_B T}{\pi\omega}\,\text{Im}\,[G_{zz}(\omega, q_{//})]_{z=0} \quad (2)$$

Here, the relationship between $I$ and $\langle|U_z|^2\rangle_{z=0}$, depends on the optical properties of the medium, the scattering geometry and the frequency and polarization of the incident light[36]. The quantity $\langle|U_z|^2(\omega, q_{//})\rangle_{z=0}$ is linked to the Projected Local Density of States (PLDOS) at the surface of the membrane, which we calculate from the out-of-plane component of the Green's function tensor, $G_{zz}$, for a given spectral and wavevector range. Further information on these calculation is available in the review by El Boudouti et al.[37].

Figure 2(a) shows high resolution spectra of the 30.7 nm thick silicon membrane. The peaks in the spectra are identified as the fundamental flexural (A0) modes of the membrane. Figure 2(b) shows the calculated PLDOS of the out-of-plane component of these modes. A single imaginary component of the frequency was included in all the calculations to fit the width of the spectral peaks due to finite phonon lifetime and instrumental broadening. The calculated spectral features are seen to follow a similar trend as the peaks in the observed spectra. This trend is related to the transformation of the modes from mainly out-of-plane polarization for small q$_{//}$ to a mixed polarization modes. To illustrate the relative magnitude of the out-of-plane density of states of this mode, a spectrum of the 30.7 nm membrane for a larger free spectral range is shown in Figure 3(a), where both the fundamental flexural and dilatational modes are observed. The intensity of the peak assigned to the fundamental flexural (A0) mode is measured to be more than two orders of magnitude more intense than that of the fundamental dilatational (S0) mode. This observation is in agreement with the Green's function simulation (Figure 3(b)), which shows a predominantly out-of-plane polarisation for the flexural mode. This is also consistent with the calculated quadratic and linear dispersion relations of the flexural and dilatational modes, respectively.

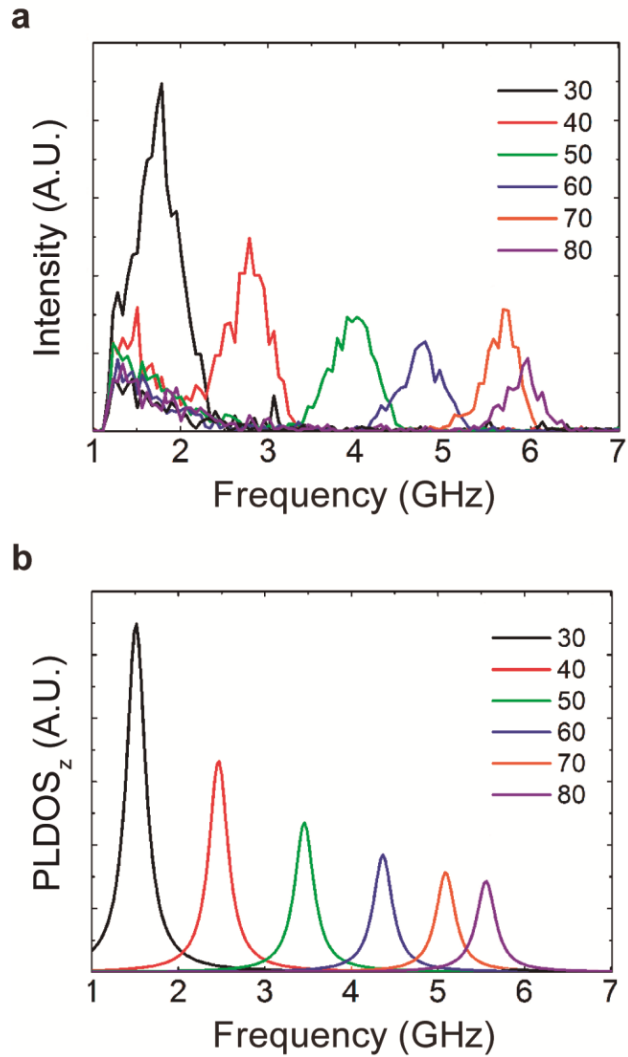

**Figure 2.** Spectra of the fundamental flexural (A0) mode of the 30.7 nm membrane as a function of the angle of incidence. (a) The ILS spectra are shown for incident angles ranging from 30 to 80 degrees with respect to the sample normal as shown in Figure 1. The measurements are performed with a resolution of ~100 MHz. (b) Green's function calculations of the projected local density of states of the out-of-plane component of the displacement, $PLDOS_z$. The ILS scattering efficiency is well described by this term due to the dominance of the ripple scattering mechanism over the photoelastic scattering mechanism as a result of the small scattering volume[37].

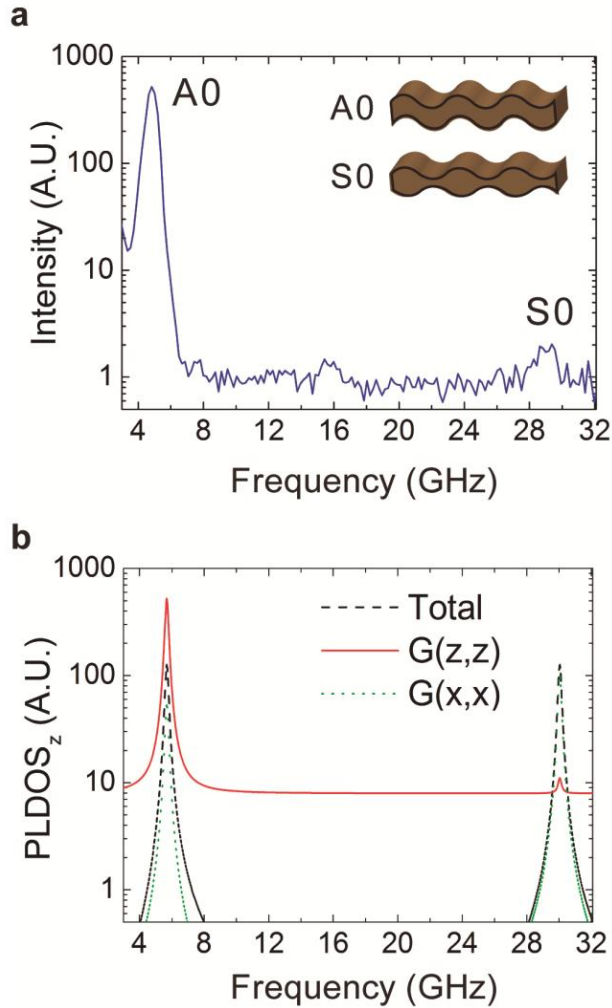

**Figure 3.** Comparison of the fundamental flexural (A0) and dilatational (S0) modes in the 30.7 nm thick membrane. (a) Inelastic light scattering spectrum recorded at an incident angle of 70 degrees ($q_{//}$ = 22 µm$^{-1}$) with a resolution of ~500 MHz. Note the intensity log scale. The inset shows schematics of the displacements of the fundamental flexural (A0) and dilatational (S0) modes. (b) Corresponding Green's function simulations showing the projected local density of states for both the in-plane (green) and the out-of-plane (red) components.

To understand the in-plane propagation of these confined acoustic modes, we compare the experimental data to dispersion relations calculated with an anisotropic continuum elasticity model, as described in the Supporting Information. The dispersion of the fundamental flexural mode for the ultra-thin membranes of various thickness values is shown in Figure 4. The calculated dispersions, with no adjustable parameters, follow closely the experimental data. The most striking features are the quadratic dispersion and the reduction of phonon frequency of the flexural mode with decreasing thickness of the membranes: $\omega = A\, d\, q_{//}^2$, where $A$ is a proportionality constant. These results are in agreement with earlier theoretical predictions for similar ultra-thin systems[21,38–40].

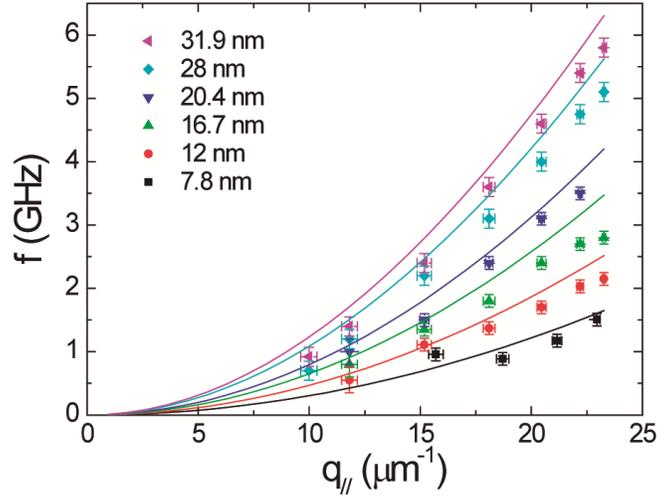

**Figure 4.** Dispersion of the fundamental flexural modes (A0). The dispersion relations are found to have a quadratic form, resulting in a larger density of states D(ω), and therefore a larger amount of energy is stored in this mode compared to linear dispersion modes. The solid lines are the calculations as described in the Supporting Information.

In addition to the remarkable drop in phase and group velocities due to the quadratic dispersion, the frequencies are consistently found to be slightly lower than those predicted by the anisotropic continuum elasticity model. While this could suggest a reduction in the effective elastic constants due to nanoscale size effects, the role of the membrane surface, including roughness and the presence of a few atomic layers of native oxide, remains to be clarified[30,41].

The effect of membrane thickness on phonon propagation is summarised in Figure 5, where the phase velocity, $v_{ph\,//} = \omega/q_{//}$, is plotted as a function of the dimensionless wavevector $q_{//} \cdot d$ for membranes with thickness values ranging from 7.8 to 400 nm. The phase velocity of the fundamental flexural mode decreases dramatically with $q_{//} \cdot d$, with a value of 300 ± 40 m s$^{-1}$ recorded for the 7.8 nm membrane (Figure 5(b)). This is more than fifteen times smaller than the surface acoustic wave velocity in the [110] direction for bulk silicon of 5085 m s$^{-1}$. As this regime corresponds to a quadratic dispersion relation, the reduction in phase velocity, $v_{ph\,//} = A\,d\,q_{//}$, is commensurate with the reduction in the group velocity, $v_{g\,//} = 2\,A\,d\,q_{//}$, down to 600 ± 80 m s$^{-1}$.

A further consequence of the quadratic dispersion associated to the fundamental flexural mode is that the density of states $D(\omega) \propto 1/d$ increases with decreasing membrane thickness, and thus more energy, $\int \hbar\omega D(\omega) f_{BE}(\omega, T)\,d\omega$, is stored in this mode, where $f_{BE}$ is the Bose-Einstein distribution function and $T$ the temperature. Therefore, the flexural mode could play a significant role in thermal transport in ultra-thin systems especially at low temperatures[42,43]. In the extreme case of graphene, this flexural mode was recently predicted to dominate the specific heat capacity and the lattice thermal conductivity[44].

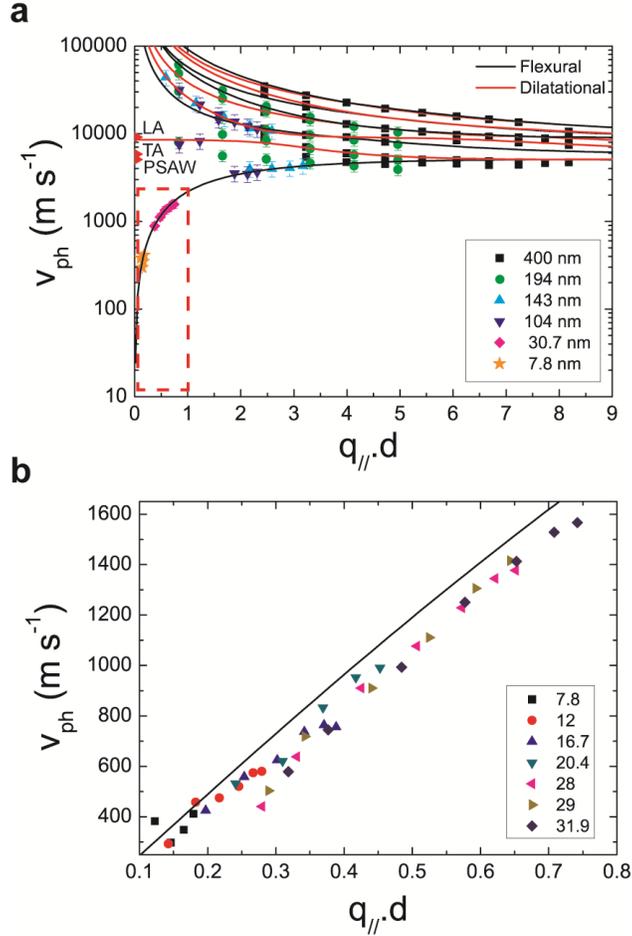

**Figure 5.** Effect of membrane thickness on phonon dispersion relations. (a) Dispersion curves plotted in terms of the phase velocity as a function of dimensionless wavevector ($q_{//} \cdot d$) for membranes with thickness values ranging from 400 nm to 7.8 nm, showing the effect of decreasing thickness. The velocities of the longitudinal acoustic (LA), transverse acoustic (TA) and Pseudo Surface Acoustic Wave (PSAW) of bulk silicon in the [110] direction are marked for reference (▶). A sharp drop in velocity is seen for the fundamental flexural mode for the ultra-thin membranes, shown by the data points enclosed by the dashed box. (b), Magnified image of the highlighted region in (a) showing data for membranes of thickness from 7.8 to 31.9 nm. The linear relationship observed is a direct result of the quadratic dispersion relation. A phase velocity down to approximately $300 \pm 40$ m s$^{-1}$ is recorded for the 7.8 nm membrane.

In summary, the in-plane propagation of confined acoustic modes in ultra-thin silicon membranes has been investigated, with measured thickness values down to 7.8 nm, in agreement with a parameter-free elasticity model. The fundamental flexural mode, which exhibits an out-of-plane polarisation and quadratic dispersion, was observed to have a scattering intensity nearly two orders of magnitude larger than the fundamental dilatational mode, which exhibits primarily an in-plane polarization and linear dispersion. Our results also show a strong reduction in the velocities of the fundamental flexural mode, in proportion to the reducing thickness. We anticipate that this membrane mode could be useful for phononic applications, such as phonon-storage, due to the strikingly-low phase and group velocities. The quantitative simulation of intensities in the ILS spectra and the quantification of the effects of modified dispersion on phonon lifetimes and thermal transport, including both the effects of the group velocity and the modified normal and Umklapp scattering processes, are subjects of current investigation.

We emphasize the need to account for the modified dispersion relation of nanostructures, in particular in the sub-20 nm regime, as the density of states, effective specific heat and thermal conductivity will possess different spectral dependencies than for bulk materials. The phonons studied here are also directly relevant for high-frequency nano-electro-mechanical systems and nano-optomechanical systems. This work provides a platform for future works of phonon engineering in nanoscale structures.

The authors acknowledge the financial support from the EU FP7 projects TAILPHOX (grant nr. 233883), NANOPOWER (grant nr. 256959), NANOPACK (grant nr. 216176) and NANOFUNCTION (grant nr. 257375); the Spanish MICINN projects ACPHIN (FIS2009-10150) nanoTHERM (CSD2010-00044), and AGAUR 2009-SGR-150 and the Academy of Finland (grant nr. 252598). The 400-nm-thick samples were fabricated using facilities from the "Integrated nano and microfabrication Clean Room" ICTS funded by MICINN. J.C. gratefully acknowledges a doctoral scholarship from the Irish Research Council for Science, Engineering and Technology (ICRSET) and E.C. acknowledges a postgraduate fellowship from the Chilean CONICYT.

**Supporting Information Available.**

- Sample fabrication and characterisation
- Brillouin Light Scattering measurements
- Dispersion relation calculations
- Green's function simulations